# How GenAI is Helping Reimagine Antenatal Care in A Low-Resource Setting: From Provider Enablement to Patient Empowerment

**Authors:** Dr. Maryam Mustafa[1], Ms. Imaan Hameed[1], Ms. Amna Shahnawaz[1], Prof. Bilal A Mateen[2,3]

**Affiliations**

1) Lahore University of Management Sciences, Pakistan
2) PATH, Seattle, USA
3) University College London, London, UK

**Corresponding Author**

Dr. Maryam Mustafa
Department of Computer Science
Lahore University of Management Sciences.
maryam_mustafa@lums.edu.pk

## Abstract

Despite steady global advances, maternal mortality remains alarmingly high in Pakistan (155 deaths per 100,000 live births in 2023); largely as a consequence of fragmented paper records, low literacy, poor access to quality healthcare, and gendered barriers that compromise care continuity. Over three years, we designed, deployed, and iteratively developed Awaaz-e-Sehat, a speech-based artificial intelligence (AI) system that generates electronic medical records (EMRs) and supports decision-making in maternal health. The tool evolved from a clinician-facing AI assistant that automated Urdu speech-to-EMR generation into a patient-centred WhatsApp-based platform, enabling women to generate their own structured clinical notes, receive AI-generated antenatal guidance, and share QR-coded records with providers anywhere in the country. This case study documents that translational journey, i.e., how the ground realities of workload, linguistic nuance, and infrastructural constraints reshaped our design. The result is not merely a new method of record-keeping, but a reimagining of antenatal care and electronic medical records themselves. In settings where clinicians are time-constrained and have little institutional incentive to document, Awaaz-e-Sehat proposes a model of care that centres patients as active participants in generating and owning their health data. By keeping patients informed about their own risk factors and integrating them into the clinical decision-support loop, the system transforms EMRs and CDSS from static institutional artefacts into dynamic tools for self-advocacy and shared accountability in maternal health.

## The Context: Fragmented Maternal Health Systems in Pakistan

Low-resource countries account for more than 90 percent of global maternal deaths; Pakistan alone experienced nearly 11,000 maternal deaths in 2023 [1,2]. Antenatal care in Pakistan is delivered through a patchwork of public, not-for-profit, and informal systems that vary widely in quality and capacity [3]. Many women first seek care from community midwives or traditional birth attendants who may lack diagnostic training and structured referral pathways [4,5]. Long distances to tertiary facilities, transportation costs, and delays in seeking or receiving emergency care further compound risk [6,7]. Clinical documentation, when it exists, is largely paper-based, comprising handwritten cards that are incomplete or inaccessible across sites, resulting in limited continuity of care between visits [8,9]. Within this fragmented ecosystem, critical symptoms are often missed, and high-risk pregnancies remain undetected until complications arise. In 2022, our team built Awaaz-e-Sehat to address this systemic fragmentation. The vision was twofold: (1) make it trivially easy to generate structured maternal-health records to support continuity of care, and (2) leverage AI to relieve overburdened clinicians while improving diagnostic accuracy. What began as an engineering challenge – turning spoken histories in regional languages recorded on simple smartphones into structured EMRs – soon became a broader experiment in redesigning the antenatal care process.

## Phase 1: A Clinician-Facing Speech-to-EMR System

The first Awaaz-e-Sehat system was designed as a modular, speech-first clinical medical record generation tool optimized for noisy, low-resource environments [10]. The proposed primary user was a clinician, and as illustrated in Figure 1, the idea was that by capturing audio responses from clinicians for a series of pre-defined questions in regional languages, that it was not only possible to generate a structured digital record, but also to overlay clinical decision support functions that would prompt the clinician to fill notable gaps and highlight red flags based on clinical guidelines.

Over the course of a seven-month (pilot) deployment (October 2023–March 2024) at a not-for-profit hospital, Awaaz-e-Sehat was used to generate 502 EMRs. Usability testing showed a high overall task completion rate of 91.44%. Among the 297 patient records evaluated in detail, the AI-generated EMRs maintained an overall field-level accuracy of 96.2%, before any editing/oversight by clinicians, who were trained to ensure the veracity of the information being stored. Error types and clinical significance are summarised in Table 1.

Field interviews showed that, beyond its technical performance, the system changed how clinicians reasoned about care, prompting more comprehensive questioning, recognition of psychosocial stressors, and attention to continuity of care between visits. Doctors described it as "a second set of eyes," particularly valuable for overburdened junior staff and nurse practitioners, navigating high patient volumes. Descriptive examples of the impact the tool had included:

- Prompting documentation that is often missing but considered part of a standard obstetric history: the assistant's miscarriage-related prompts (e.g., gestational age, whether a D&C was performed) were identified as being helpful for ensuring that critical details were consistently documented.
- Prompting documentation around nuanced issues that are specific to the patient. In one case, the family history questions prompted by the AI assistance surfaced whether a congenital condition originated on the maternal or paternal side, a question clinicians noted was not typically captured in their routine workflow. In another case, structured questioning facilitated the disclosure of domestic violence, underscoring the potential of guided prompts to support identification of psychosocial risks.

*Table 1: Frequency, Examples and Clinical Relevance of Errors Made by Awaaz-e-Sehat*

| Category | Frequency | Definition | Examples |
| --- | --- | --- | --- |
| No Action Needed | <3% | The inaccurate EMR field is insignificant and does not impact clinical decision-making, so it is disregarded. | - Spelling mistake in place of birth in "previous pregnancies".<br>- Patient's education marked as "Metric" instead of "Matric".<br>- Gravida marked "Prime (First Pregnancy)" instead of "Primigravida (First Pregnancy)". |
| Easily Identifiable and Correctable | <1% | The doctor can easily spot the inaccuracy by reviewing the patient's complete EMR and make the necessary correction without additional input. | - Incorrect medicine name (Duplascon instead of Duphaston).<br>- "Vaginal bleeding since last period" marked "yes since a week ago" instead of "since 23rd August 2023".<br>- Family status marked as unknown instead of nuclear family. |
| Unidentifiable and Uncorrectable Without Ground Truth | <2.5% | The errors cannot be identified or corrected by the doctors without access to the original recording or an external ground-truth EMR. | - Age of baby in "previous pregnancy details" marked as 3 years instead of "No Info".<br>- Spelling mistake in husband's name.<br>- "Abortions" marked as 2 instead of 1. |

Furthermore, the system's potential clinical impact is evident in its ability to generate red-flag alerts for high-risk cases. In total, red-flag alerts were generated for 163 of the 297 aforementioned patients' records analysed in detail, amounting to 335 distinct alerts. Medical accuracy and patient-specific relevance were assessed as binary variables, with consultant obstetricians/gynaecologists rating each red-flag alert as either accurate/inaccurate and relevant/not relevant. Across all reviews, 96.3% of alerts were judged medically accurate, and 93.9% were judged relevant to the specific patient case. Unfortunately, the vast majority were not acted upon – not because clinicians doubted their accuracy, but because sustained follow-up required time, coordination, and systemic continuity that the workflow did not support. Similar issues of providers not acting on 'useful' insights have been observed in other studies evaluating the impact of GenAI-based clinical decision support systems [11,12]. A comprehensive description of the user experience and technical validation research conducted based on the pilot data is provided elsewhere [10].

**Figure 1.**

**Clinical Conversation Capture:**
Using a smartphone's microphone for voice capture, clinicians were able to engage in their standard clinical history taking, but instead of typing, a fine-tuned multilingual speech-to-text solution (based on the Whisper model) transcribed and translated the conversation in preparation for populating an electronic medical record.

**LLM-driven EMR Generation:**
GPT-4 was prompt-engineered to:
(1) generate structured EMRs from the transcripts, (2) identify clarifying follow-up questions for ambiguous entries (e.g., severity grading of an episiotomy), and
(3) cross-check consistency with a bilingual medical dictionary curated from real-world data (e.g., "sugar" = "diabetes").

**Red Flag Identification:**
Systematic comparison of the digitised information against known risk indicators from relevant guidelines — such as elevated blood pressure, anaemia markers,or specific obstetric history patterns — was presented as real-time alerts for clinicians within the EMR.

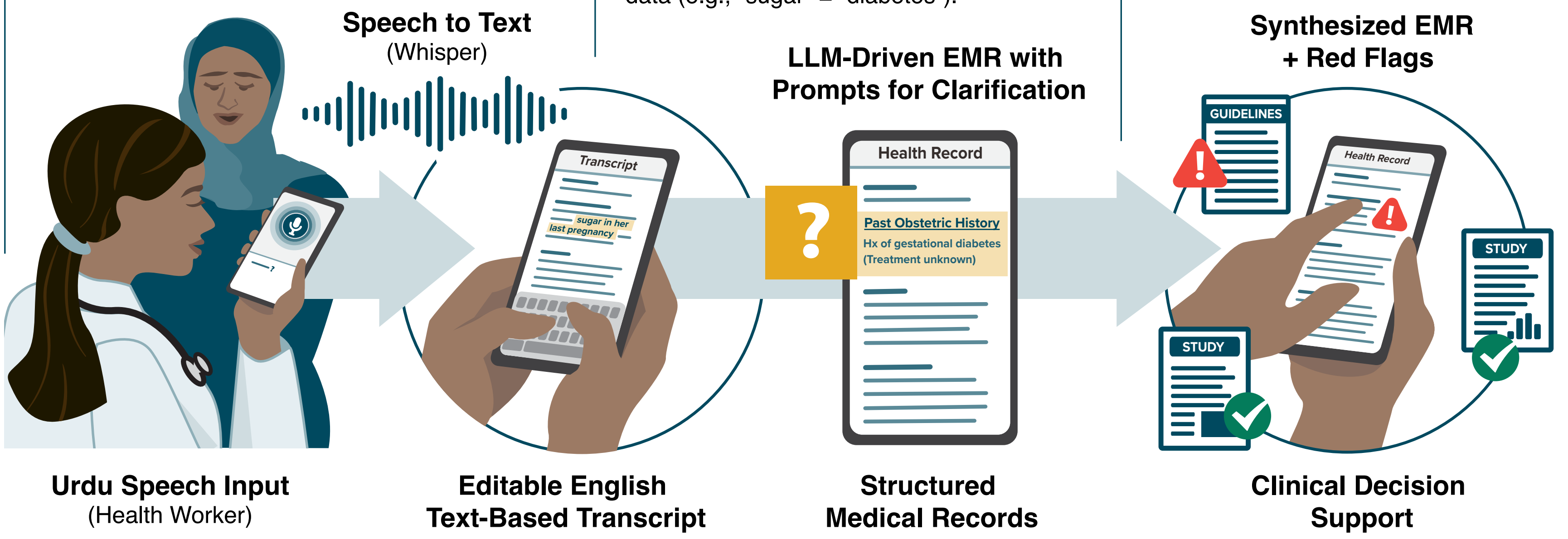

**Figure 1: The original provider-facing Awaaz-e-Sehat workflow.**
An LLM-assisted pipeline for capturing Urdu clinical conversations, to generate structured electronic medical records, and surface guideline-based red-flag alerts to support clinical decision-making. The system's architecture combined structured medical logic with locally grounded linguistic norms. For example, the EMR template was co-developed with obstetricians and mid-career residents, aligning with WHO and RCOG antenatal parameters, but reorganised to mirror how doctors actually gathered histories in real time. Iterative fieldwork revealed that clinicians preferred non-linear entry and conversational flow, leading to an interface that allowed flexible navigation rather than rigid form completion. Moreover, to ensure local relevance, the multilingual Whisper-large-v2 model was fine-tuned on locally recorded hospital speech to handle Urdu-English code-switching, variable accents, and background noise. The fine-tuned model achieved a 16.3% word-error rate – nearly a three-fold improvement over baseline multilingual models – enabling accurate transcription even amid clinical interruptions.

## Transitioning From Clinician Workflows to Patient Realities

Despite the tool's potential impact, field deployment revealed deep structural constraints that the provider-facing technology alone could not overcome [10]. Consultations remained brief and crowded, leaving little time to capture full clinical histories. Connectivity fluctuations interrupted uploads, and the very staff who most benefited from automated documentation were those least able to sustain its use [13,14]. More fundamentally, there was limited will or institutional incentive for clinicians to take on additional documentation tasks – regardless of how easy, intuitive and seamless the tech was. Record-keeping was perceived as administrative work, peripheral to the diagnostic encounter, and disconnected from any formal reward or accountability structure. The latter is related to the fact that many doctors were seeing large numbers of transient patients who might not return for subsequent visits; thus, there was no institutional mechanism and little personal motivation to act on identified risks or to track outcomes once a consultation ended. In essence, whilst ambient scribes (to which this tool is analogous) are increasingly being seen as a valuable solution for addressing provider documentation burden in high-income countries [15], this is clearly not a ubiquitous experience. In Pakistan, digital interventions often succeed or fail not because of technical capability but due to entrenched social norms that limit women's autonomy and technology use [16].

Amid these limitations, one insight emerged consistently across hundreds of recorded consultations: patients themselves were the only constant in this fragmented system. Prior research has shown that women often use digital channels cautiously due to privacy concerns, phone sharing, and limited control over technology in households [16]. The same women who recounted detailed symptoms, family histories, and contextual challenges during clinic visits were also the most persistent source of health information across encounters and sites. The problem, then, was not a lack of data but where, and by whom, it was entered into 'the system'. Recognizing that clinicians operated under structural constraints while patients already held and articulated the relevant information reframed our design question entirely: how might antenatal records be generated in contexts where clinicians are time-constrained and patients already narrate their own experiences across multiple maternal health workers and sites during their pregnancies? It also highlighted the question of who should own patients' medical data - hospitals/clinics where clinicians digitally record it or patients themselves? In the current system of physical paper records, patients own their medical data in physical files that they must keep track of and carry to each ANC visit. The pilot reframed our understanding of where maternal health data originates and who sustains it, and helped us question the issue of ownership of digital health records. In a system where clinicians are transient actors, but patients remain the only constant, it became evident that continuity must begin with them. Re-centring patients as the primary data stewards could both ease pressure on doctors, who no longer had to spend scarce consultation time entering histories, and give women more control by allowing them to actively share their information and better understand their own records.

## Phase 2: A Patient-Centered EMR Generation Workflow

Through iterative co-design sessions with obstetricians and patients, we converted standard antenatal visit clerking questions into short, culturally sensitive WhatsApp exchanges that were suitable for collecting information directly from pregnant women. WhatsApp was selected because it is the most widely used messaging platform among the communities in question. In this system (as shown in Figure 2), brief WhatsApp-based interactions including text, voice notes, and reports (via images) are captured and organized into a structured electronic medical record that can be forwarded to clinicians for review (via QR code) and follow-up. Alongside data collection, users can converse with the system about symptoms and receive tailored antenatal guidance, feedback, and reminders, while maintaining oversight of their own medical history. This design centralizes antenatal information in a single, patient-controlled record, reduces the clerical burden on clinicians, while ensuring that all AI-mediated interactions are verified within clinical workflows. One of our biggest initial challenges was the length of interactions. While doctors can collect histories conversationally over a prolonged period, creating a comprehensive digital equivalent risked burdening patients, who vary in comfort with technology and literacy required more translation. This is a particularly nuanced challenge for our context given that patients often lie about data points they feel reflect poorly on them [17]. Clinicians handle this in person by asking the same question in different ways and estimate the truth through multiple answers to different questions. This becomes painstakingly long and complicated to replicate in digital communication. These challenges echo broader findings that many Pakistani women rely on simple, familiar digital tools due to constraints in literacy, device access, and gatekeeping [18,19]. We therefore spent considerable time iterating with clinicians to balance the need to capture detailed antenatal histories with the practical need to reduce the time and effort required of women to respond.

The result was a streamlined set of initial questions, with a cascading flow that prompted additional questions when certain answers were provided, with the expectation that this would be done over several encounters rather than in a single setting. As such, each message was designed to solicit information about a specific parameter (e.g., blood pressure, fetal movement, family history) in everyday Urdu, and an LLM was used to parse responses—often code-mixed Urdu-English or Roman Urdu—to populate structured EMRs. Several user-research-derived insights informed the final design of the system with the intent of maximising the likelihood of uptake, engagement and retention, including:

1) Sensitive topics were prefaced with gentle framing (”Maaf kijiye ga…” [please excuse me] before smoking or substance-use questions).

2) To accommodate low literacy, the system was built to support multiple response modalities – text, numerals, or voice notes – and questions were framed accordingly.

3) To ensure the system could interpret how women naturally describe their symptoms, we developed a customized bilingual dictionary that mapped common colloquialisms, euphemisms, and metaphors about reproductive and maternal health to their corresponding clinical terms. Expressions such as ”neeche wali jaga” [English: ‘the area below’] for vagina, ”charbi” [English: ‘fat’] for discharge, and ”haddi pighal rahi hai” [English: ‘bone is melting’] for infection were aligned with standardized medical vocabulary, allowing the LLM to accurately parse and structure narratives into electronic medical records. This layer of linguistic grounding reduced errors in symptom interpretation and ensured that patient speech, however informal, could be meaningfully transformed into medically coherent data. More details on this nuanced use of language and the dictionary development are described elsewhere [20].

**Figure 2.**

**Patient-Facing Maternity Assistant:**
Using the WhatsApp user interface, patients interact with an LLM-powered virtual health assistant that collects comprehensive obstetric histories via conversational prompts, including voice notes to address literacy barriers. The system identifies missing or critical information, follows up with targeted questions, and allows patients to ask their own questions, generating personalized responses grounded in their medical record. Histories are gathered over multiple interactions to maintain engagement and improve data quality, and the outputs are converted into structured electronic medical records and securely stored in the cloud.

**Provider-Facing EMR & CDSS:**
Patients securely share their medical records with clinicians via a QR code, providing instant access to an up-to-date obstetric history and thus reducing time spent on intake. A clinical decision support engine flags key risks, helping providers focus on the most critical issues during the limited consultation time. Clinicians can update records during visits, and patients can later add new information — such as test results from other providers — keeping the record current across encounters. After appointments, the virtual doctor follows up to explain red flags as well as any test results, and to provide relevant reminders in clear, patient-friendly language.

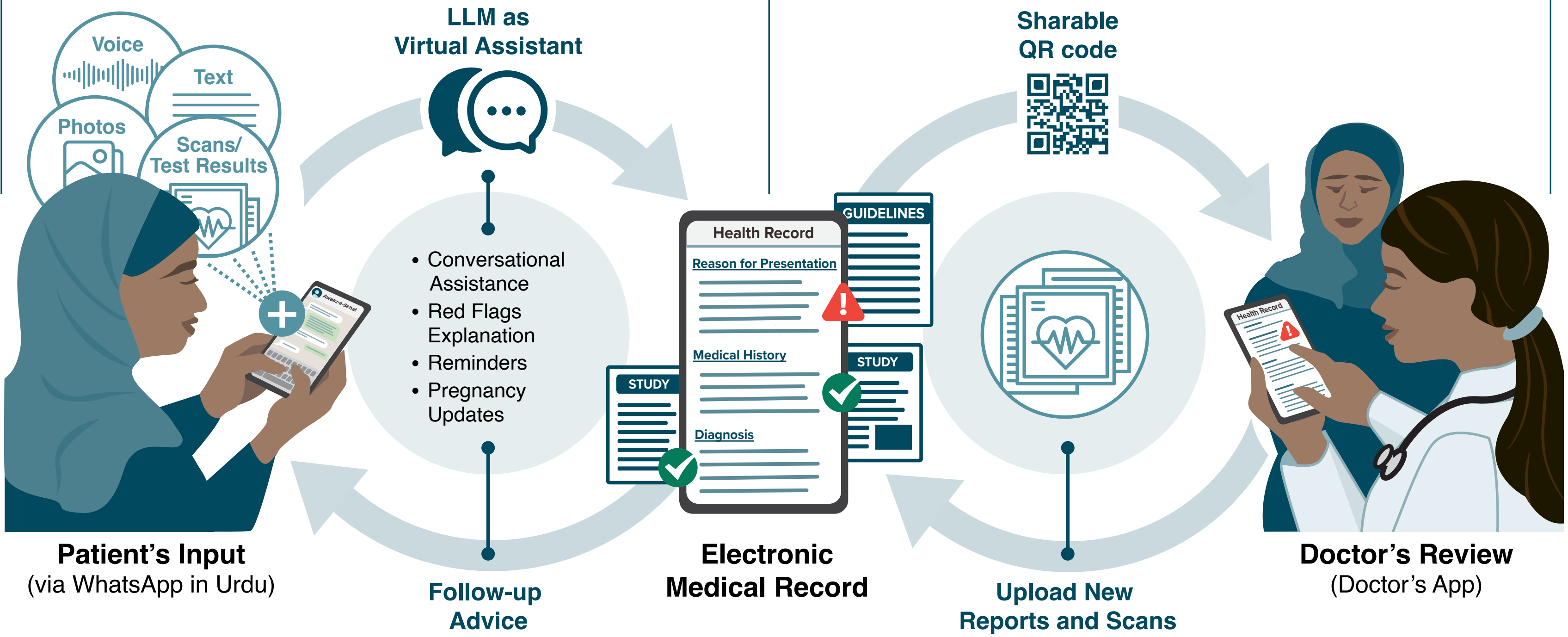

**Figure 2: The revised patient–clinician Awaaz-e-Sehat workflow.**
WhatsApp-based Urdu inputs (voice, images, and reports) are processed by an LLM to generate and validate electronic medical records, while enabling patients to ask about symptoms and receive tailored advice, and supporting clinician review and follow-up through a shared digital record.

To enable information exchange, we built a mechanism for patient-controlled record portability directly within WhatsApp; each woman can generate a QR code for the EMR generated via her interactions with the Awaaz-e-Sehat system, which she can show at any facility. With her consent, clinicians can scan the code to access prior histories, lab results, and ultrasound reports. New information added by one provider automatically updates the record for future visits, allowing care to continue seamlessly across sites. This design transforms what was once a static, paper-based artefact into a living digital record owned by, primarily generated by, and carried by the patient herself. In other words, it creates continuity in a system that has historically lacked institutional memory, whilst enabling women to be custodians of their own medical history. Whilst many have previously suggested ledger-based distributed systems for devolving health information management back to the true data owners (i.e., patients) [21,22], in this work, we have created a fit-for-context variation that achieves the same principled outcome.

## Phase 3: From Information System to Comprehensive Antenatal Health Assistant

In parallel to the transition from provider-facing to patient-facing tool, and as a result of direct engagement with pregnant women, a new demand signal emerged. Women began approaching the field-testing team with pregnancy-related questions – about symptoms, nutrition, medications, and test results – reflecting a more profound absence of accessible, trustworthy antenatal information. Most had no reliable source of guidance outside clinical visits, and digital search tools such as Google were inaccessible to those who could not read English or discern accurate information from misinformation. Prior research has shown that women often use digital channels cautiously due to privacy concerns, phone sharing, and limited control over technology in households [23]. In practice, this meant that for many women, learning about their own pregnancy depended entirely on brief, hurried clinical encounters with busy physicians. Addressing this gap required rethinking Awaaz-e-Sehat, moving from a data-collection system to a directional channel that facilitated information capture whilst providing a safe and reliable means for women to seek and receive information in their own language, without depending on literacy.

To achieve the above, we extended the system into a conversational pregnancy assistant that allowed women to ask questions directly in their local language through speech. The assistant was guided by a carefully designed prompt that established the clinical context, communication norms, and tone, ensuring it responded in familiar Urdu. Its medical reasoning was grounded in the Royal College of Obstetricians and Gynaecologists (RCOG) guidelines, and we supplemented this with an internally developed FAQ document informed by consultations with our partnering Obstetrics & Gynaecology departments. This FAQ provided examples of the most commonly asked questions we had encountered in early testing, allowing the model to better handle specific, real-world concerns.

Beyond answering questions, the assistant was designed to actively educate and empower women throughout their pregnancy; for example, it sends routine informational updates that were structured around four domains: nutrition guidance relevant to locally accessible foods, explanations of fetal growth and development, awareness and mitigation of common risk factors and support related to the mother's emotional wellbeing. All of these updates were written in simple, conversational Urdu that mirrors how clinicians speak to patients (Figure 3). Each message underwent medical vetting to ensure accuracy and to strike an appropriate balance between clarity and clinical responsibility, avoiding both overwhelming detail and overly simplistic advice. By combining responsiveness with these proactive, stage-appropriate updates, the virtual pregnancy assistant becomes better placed to be both a source of reassurance for immediate concerns and an ongoing companion throughout the antenatal journey.

Finally, we sought to redesign the tool's clinical decision support component. In the clinician-facing prototype, the LLM-powered red-flag identification tool served solely as a nudge visible to doctors. In the redesigned platform, these alerts were reimagined to inform and empower women directly. Each warning was translated into clear, conversational Urdu – explaining what the risk meant and why follow-up mattered (Figure 4). The intention was not just to notify but to equip women with the language and confidence to advocate for themselves; to request tests, to seek clarification, and to insist on care when something felt wrong. In a health system as fragmented and uneven as Pakistan's, where continuity is not guaranteed and referrals are unreliable, enabling women to recognise and articulate their own risk became a form of structural resilience.

## Acknowledging the Elephant in the Room

If we'd done this under the jurisdiction of the EMA or the UK MHRA, there is every likelihood that this solution could have been labelled a 'medical device' and thus subject to regulatory oversight [24]. In Pakistan, where the national regulatory body is currently focused solely on medicines and vaccines, innovators are forced to work in a grey area. The challenge therein is that academics and social impact-focused entrepreneurs are asked to shoulder a non-trivial risk burden or steer clear of innovating in this way altogether. The latter is untenable as it means continuing to accept substandard care in women's health that contributes to the profound mortality risks prospective mothers face; thus, more investment is needed to facilitate AI-focused innovation in healthcare settings that enable collaboration with regulators and the development of adequate guardrails. High-income countries have begun launching regulatory sandboxes in service of these aims [25,26], and pioneering middle-income nations like South Africa are following suit [27]. Still, opportunities for the rest of the world remain few and far between.

## Conclusion: Reimagining Antenatal Care for Fragmented Health Systems

In overburdened health systems defined by fragmentation and profound variations in quality, AI cannot afford to be naïve. The path to equity does not lie in layering broken processes with new technologies but in confronting why they fail and designing for those failures from the start. The piloting, reimagining, and scaling of Awaaz-e-Sehat show us that the future of antenatal care cannot rely on idealized assumptions of diligence or continuity. Instead, it must redistribute the work of care toward those with the most consistent stake in it: patients themselves. When women generate their own data, understand their risks, and carry their records, AI becomes an instrument of participation and resilience rather than substitution. This reframing transforms AI-based decision support from a unidirectional extraction of information, translated into insight by a clinician, into a multifaceted dialogue spanning patient-AI, AI-clinician, and clinician-patient interactions, where, critically, knowledge and responsibility are shared.

**Figure 3.**

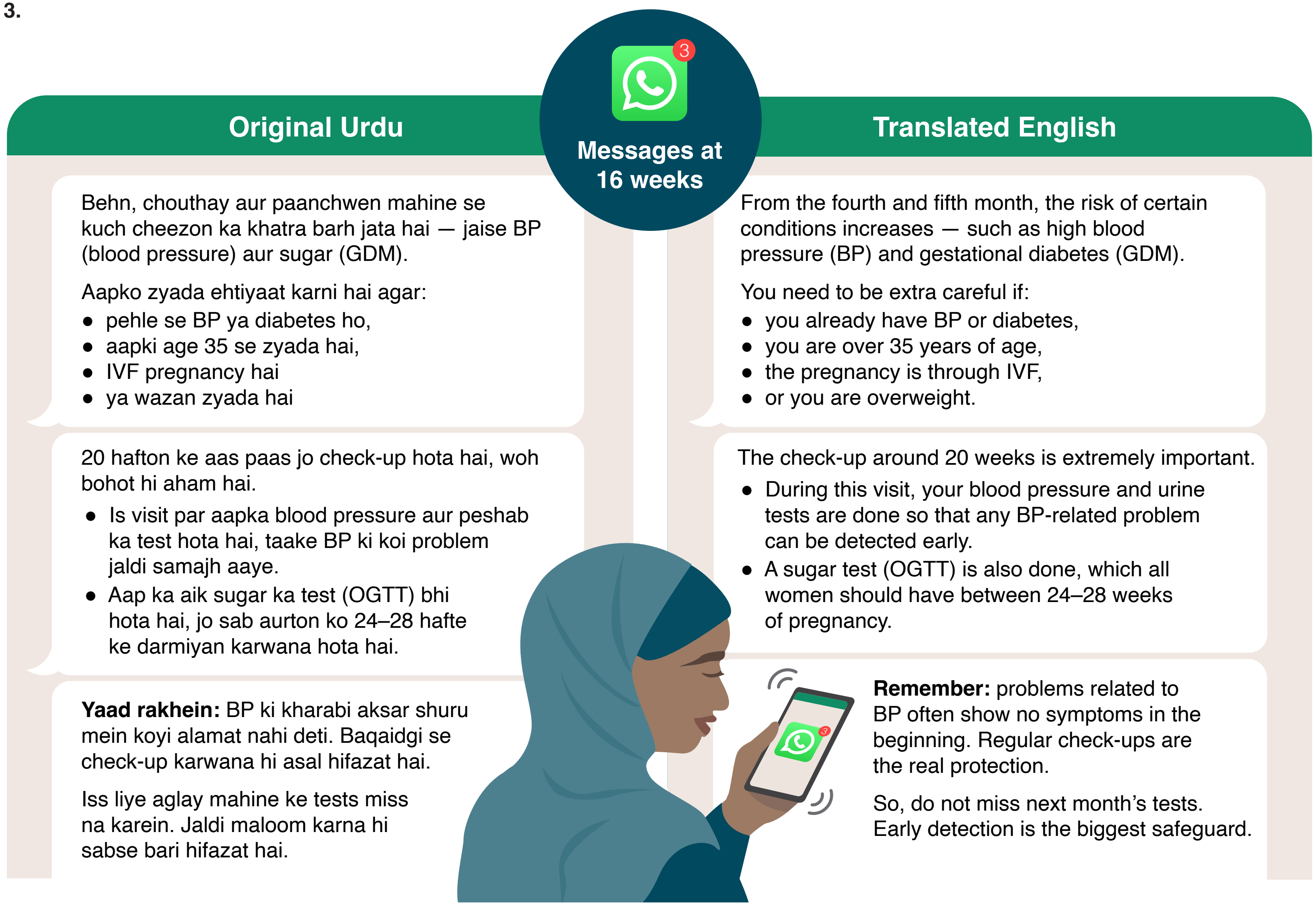

*Figure 3: Example of a Reminder Provided by the Revised Awaaz-e-Sehat System*

Figure 4.

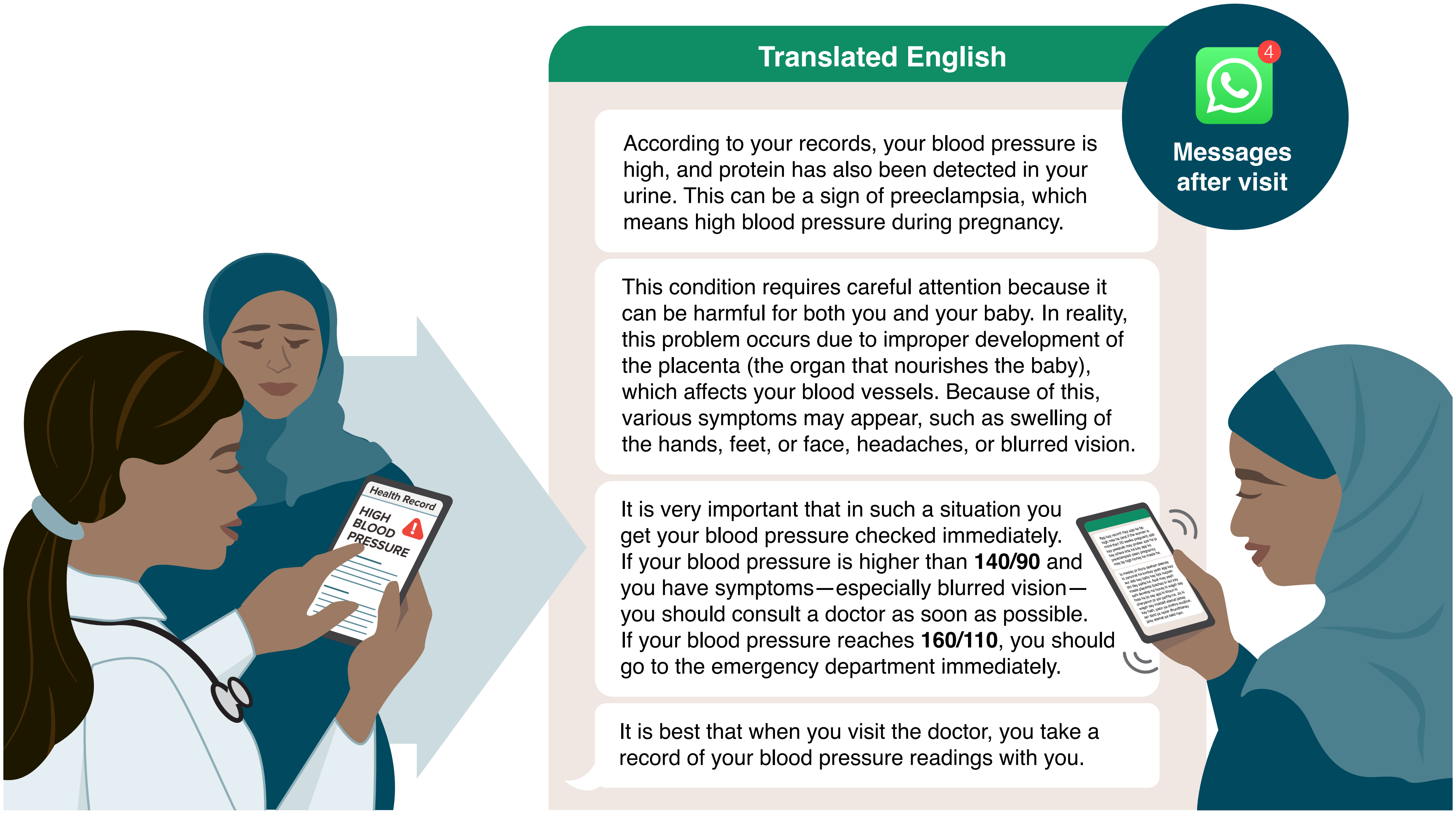

*Figure 4: System Output in response to Increased Blood Pressure/Possible Preeclampsia Risk*

# Other Information

**Funding**

The work described in this case study was funded by the Gates Foundation (INV-068056), and Patrick J. McGovern Foundation (1319). The funder(s) did not influence study design, data collection, analysis, interpretation, or the decision to publish the findings.

**Data Availability**

There is no data associated with this case study.

**Code Availability**

There is no code associated with this case study.

**Acknowledgments**

We acknowledge the support provided by Shalimar Hospital, Dr. Beena Ahmed at UNSW and Dr. Fozia Umber Qureshi.

**Author contributions (CRediT taxonomy).**

Conceptualization: MM & BAM. Supervision: MM. Funding acquisition: MM & BAM. Writing (original draft): All Authors. Writing (review and editing): All Authors.

**Competing Interests**

The original Awaaz-e-Sehat system was developed with philanthropic grant funding, and the intellectual property was eventually spun out by the first author (Maryam Mustafa) into a social-impact not-for-profit organisation, bound by global access guarantees. See Gates Foundation Statement on global access commitments: https://www.gatesfoundation.org/about/policies-and-resources/global-access-statement),

**Ethics Approval**

Ethics approval was not required for this narrative case study. The underlying research studies that informed the case study each had its own ethics approval from the appropriate IRB.